# Study of Supernovae Important for Cosmology

**P. V. Baklanov**[a, b], **S. I. Blinnikov**[a−c], **M. Sh. Potashov**[b], and **A. D. Dolgov**[a, b, d]

[a] *Institute for Theoretical and Experimental Physics, ul. Bol'shaya Cheremushkinskaya 25, Moscow, 117218 Russia*
*e-mail: baklanov@gmail.com, sergei.blinnikov@itep.ru*
[b] *Novosibirsk State University, ul. Pirogova 2, Novosibirsk, 630090 Russia*
*e-mail: marat.potashov@gmail.com*
[c] *Sternberg Astronomical Institute, Moscow State University, Universitetskii pr. 13, Moscow, 119992 Russia*
[d] *University of Ferrara and INFN, 44100 Ferrara, Italy*
*e-mail: dolgov@fe.infn.it*

The dense shell method for the determination of distances to type-IIn supernovae has been briefly reviewed. Applying our method to SN 2006gy, SN 2009ip, and SN 2010jl supernovae, we have obtained distances in excellent agreement with the previously known distances to the parent galaxies. The dense shell method is based on the radiation hydrodynamic model of a supernova. The method of the blackbody model, as well as the correctness of its application for simple estimates of distances from observation data, has been justified.

## 1. DENSE SHELL METHOD AS A METHOD FOR THE DETERMINATION OF DISTANCES TO SUPERNOVAE

Supernovae belong to the brightest astronomical objects. For this reason, they play a very important role in the measurements of distances in the Universe and in verifications of cosmological models. Type-Ia supernovae are observed up to redshift $z = 1.7$. Three type-IIn(GH IIn) supernovae at $z \approx 0.8$, 2.0, and 2.9 were identified by spectra.

Owing to type-Ia supernovae, it was revealed that the Universe expands now with acceleration [1−3]. Within the Friedmann model, this can be interpreted as the presence of a nonzero cosmological term. This discovery was awarded the Nobel Prize in physics in 2011.

In a more general class of the models, this result can be interpreted as the presence of dark energy in the Universe, which is responsible for gravitational repulsion at cosmological distances. At present, one of the most important problems of fundamental physics is the determination of the reality and properties of dark energy (and dark matter). Supernovae observed at cosmological distances will continue to play a key role in the solution of this problem.

Several methods for the use of supernovae and their gaseous remains in cosmography were proposed. They can be classified into two groups. The methods of the first group involve the idea of a standard candle. It was previously thought that type-Ia supernovae are standard candles; i.e., the maxima of their absolute luminosity (i.e., light power) in different bursts are identical. More recently, it was found that this is not the case [4]. However, procedures that allow the determination of the absolute luminosity, i.e., the standardization of a candle, were proposed [5].

Nevertheless, many factors that can affect the results obtained for cosmology with the use of Ia supernovae remain. In particular, these are intergalactic absorption, redshift in parent galaxies (see [6, 7] and references therein), change in the metallicity of the progenitors of supernovae, and change in a relative role of different supernovae at different ages of the Universe.

Another possible source of errors is an incorrect classification and admixture of unusual events such as type-Ia supernovae. For example, a strange class of type-Ia supernovae, the SN 2002cx subtype of supernovae, was discovered. They are weak, but slow (see, e.g., [8]); i.e., their behavior is opposite to the Pskovskii−Phillips relation, which is used in cosmology. According to the Pskovskii−Phillips relation, slowly decreasing type-Ia supernovae are the brightest. Let the number of SN 2002cx subtype events increase with the cosmological redshift $z$. Using the Pskovskii−Phillips relation, which is established for nearby type-Ia supernovae (i.e., at $z = 0$), we conclude that type-Ia supernovae at large $z$ values seem on average *weaker*. Consequently, the photometric distance to them is longer than that at the actual $\Omega_\Lambda$ value. As a result, a false contribution to dark energy would be obtained.

Conley et al. wrote [9]: "In the absence of a physical model, it will always be possible to imagine an evolutionary scenario which will pass all available tests, yet which will bias the cosmology by an arbitrary amount." This does not mean that supernovae cannot be used for reliable cosmography. It is necessary to develop new approaches to this problem. One of such approaches is discussed in this brief review.

"Standard candle" methods require distances to a large number of supernovae that are measured by another independent method involving the cosmic distance ladder [10]. Without large statistics of objects with known distances, it is impossible to calibrate the candle standardization procedure (see, e.g., reviews [11, 12]). Thus, supernovae are used in this group of methods as secondary distance indicators.

Our method belongs to another group of methods that use supernovae as primary distance indicators. Type II supernovae are generally strongly different in luminosity and in shape of the light curve and cannot be used in the standard candle method. At the same time, their important advantage is the possibility of direct measurements of distances by, e.g., the expanding photosphere method. This method does not require the candle standardization procedure and cosmic distance ladder.

High-quality spectral observations of supernovae make it possible to determine distances by the spectral fitting expanding atmosphere method [13]. In contrast to the expanding photosphere method, this method does not require the assumption of the blackbody spectrum of a supernova.

In this work, we review the development of a new method for the measurements of cosmic distances which is partially based on the expanding photosphere method, spectral fitting expanding atmosphere method, and expanding shock front method [14]. This method should involve the brightest type-IIn supernovae, which have been actively discovered and studied in detail in recent years [15].

We propose to refer to this method as the dense shell method because the luminosity of type-IIn supernovae is determined by the propagation of a thin dense layer in the environment. It is shown by examples of SN 2006gp and SN 2009ip that the dense shell method is relevant: the distances to these supernovae were obtained independently of usual calibrations of distances. This new method does not require the standard candle approximation, as for type-Ia supernovae, or the cosmic distance scale (ladder).

Photons in type-II supernovae are created in shock waves propagating in their shells (in time $<10^4$ s in SN 1987A and in up to $10^7$ s or longer in type-II supernovae). A shock wave in usual type-II supernovae creates not only a short-term burst of hard radiation but also a reservoir of entropy ensuring luminescence in the "plateau" stage for several months. In type-IIn supernovae, where the shock wave propagates in the environment for several months, it is a source of luminescence [16–19].

The light curves of type-II supernovae are very diverse in shape and amplitude. Consequently, they cannot be described by the standard candle, i.e., a certain unified light curve. The light curve strongly depends on the properties of the shell surrounding the source of the energy of a supernova, such as a collapsing core or nuclear fusion in the core. At the same time, the shell makes a type-II supernova much less dependent on the details of an explosion. A real photosphere is observed in the shell for several months and is manifested in light curves in the form of a classical plateau.

The idea of the expanding photosphere method abbreviated as EPM goes back to Baade [20] and Wesselink [21], who developed it to determine distances to variable stars, cepheids.

Since we can develop a detailed model of type-IIn supernovae, a new direct method, dense shell method, can be developed to use the bright light of type-IIn supernovae in cosmology.

The essence of this method is as follows. The radius of the photosphere moving at the velocity $v$ changes in the measurement time $dt$ by $dr = vdt$. Thus, when the $v$ value is known, change in the radius $dr$ can be obtained without any cosmic distance ladder. The measured radiation flux is given by the expression

$$F = 4\pi r^2 \sigma T^4/D^2, \qquad (1)$$

where $D$ is the photometric distance, which does not change, and $T$ is the temperature, which can be measured. The $dr$ and $dF$ values are measurable. It is more convenient to use the quantity $S = \sqrt{F}$, i.e.,

$$S = 2\sqrt{\pi\sigma} r T^2/D. \qquad (2)$$

If $T$ remains almost unchanged between two measurements,

$$dS = 2\sqrt{\pi\sigma} dr T^2/D. \qquad (3)$$

The $dr$ value can sometimes be measured directly: $dr = v_{ph}dt$, where $v_{ph}$ is the velocity of the photosphere. Then, the distance $D$ is determined directly in terms of the measured $dS$, $dr$, and $T$ values.

The authors of [22] clearly understood that the velocity determined from spectral lines is the velocity of matter $u$, whereas the photosphere moves with respect to matter (the absorption coefficient of matter decreases in the process of expansion). Even the signs of $u$ and $v_{ph}$ can be opposite when the photosphere contracts. This is the main difficulty of the expanding photosphere method and spectral fitting expanding atmosphere method: these methods are applicable under the assumption that free expansion occurs and the velocity of matter is $u = r/t$. This is the case when dense matter is absent in the environment at a certain time after the explosion. At the same time, type-IIn supernovae are surrounded by dense matter and the shock wave does not reach a rarefied medium for several months or even years.

On the other hand, according to plots presented in [19, 16], the entire matter behind the shock front in such supernovae is compressed into a cold dense shell. The photosphere is "glued" to this dense shell. As a result, $u = v_{ph}$ and can be measured. All above properties correspond to Baade's idea [20] proposed in the 1920s.

Finally, we can formulate the new dense shell method for the determination of cosmological distances with the use of type-IIn supernovae. This method includes the following stages.

—The *narrow* components of spectral lines are measured to estimate the properties (density, velocity) of the near-star shell. This stage does not require a high accuracy of measurements and simulation.

—The *broad* emission components of lines are measured and the velocity at the level of the photosphere is found (with maximum possible accuracy).

—Although the law is $u = r/t$ is inapplicable in these supernovae, $v_{ph}$ now corresponds to the *real* velocity of the radius of the photosphere (rather than only to the matter ejection rate, as in type-II-P supernovae).

—Baade's initial idea [20] of the measurement of radius increments $dr = v_{ph}dt$ by *integration with respect to time* is now used (certainly, with necessary allowance for scattering, darkening/brightening toward the edge, etc.). The resulting radius values should be used in iterations of the optimal mode.

—The distance $D$ can be obtained in terms of the observed flux by the expression

$$D = 2\sqrt{\pi\sigma}\,dr\,T^2/dS. \tag{4}$$

Such a simple approach is inapplicable when $T$ varies significantly. In this case, it is necessary to construct a model that provides the best reproduction of the observations of broadband photometry and the velocity $v_{ph}$, which is controlled by observation of $dr(t)$. Such a model is necessary for the calculation of the evolution of $r$ and the correction factor $\zeta$, as well as for detailed predictions of the theoretical flux $F_\nu$.

We present the main algorithm of the method based on the blackbody model with the temperature $T$ and correction factor $\zeta$. We assume that observations are frequent enough to measure variation of the radius $dr = v_{ph}dt$. Let $R_0$ be the initial (unknown) radius and $R_i = R_0 + \Delta R_i$ ($i = 1, 2, 3, ...$), where the $\Delta R_i$ values are known from summation of measured $dr$ values. Then,

$$\zeta_i^2(R_0 + \Delta R_i)^2 \pi B_\nu(T_{ci}) = 10^{-0.4A_\nu}D^2 f_{\nu i}, \tag{5}$$

or, taking the square root,

$$\zeta_i(R_0 + \Delta R_i)\sqrt{\pi B_\nu(T_{ci})} = 10^{-0.2A_\nu}D\sqrt{f_{\nu i}}. \tag{6}$$

A good model gives the set of $\zeta_i$ and $T_{ci}$ for all observation points. In terms of the measured $f_{\nu i}$ and $\Delta R_i$ val-ues, the unknown $R_0$ values and combination $10^{-0.4A_\nu}D^2$ are determined by the least squares method. To obtain the distance $D$, it is necessary either to know the absorption $A_\nu$ from astronomical observations or to determine it from the same equations written for various spectral filters. In this case, knowing the $R_0$ value, we obtain a series of equations of the form

$$10^{-0.4A_\nu}D^2 = a_s, \tag{7}$$

where $a_s$ is a certain constant depending on a particular UBVRI filter marked by the subscript $s$. This procedure gives the difference $A_{s1} - A_{s2}$. Then, $A_\nu$ can be obtained from the frequency dependence of extinction.

## 2. APPLICATION OF THE DENSE SHELL METHOD TO PARTICULAR EXAMPLES

Using the dense shell method described above, we obtain the distances to three type-IIn supernovae, SN 2006gy, SN 2009ip, and SN 2010jl.

### 2.1. SN 2006gy

The SN 2006gy supernova, the brightest of all known supernovae, was discovered on September 18, 2006 (UT), near the center of the NGC 1260 galaxy in the framework of the Texas Supernova Search program [23]. The total absorption to the supernova is determined with some spread (see discussion in [24]). We used the value $A_R = 1.3 \pm 0.25^m$.

In [25], using the observation data [26–28], we estimated the distance to the SN 2006gy supernova as $D \approx 69^{+19}_{-15}$ Mpc. This distance is in good agreement with the known distance modulus to the parent galaxy $\mu = 34.53^m$ [28], which is equivalent to $D = 71$ Mpc.

### 2.2. SN 2009ip

The remarkable SN 2006ip supernova, located in the NGC 7259 galaxy after its discovery in 2009 [29], produced several ejections analyzed in detail by observers.

The brightest burst was manifested as a rapid growth after September 24, 2012, owing to the interaction of the rapidly propagating ejection with the circumstellar medium [30]. Unfortunately, broad lines disappeared almost completely after September 28. This circumstance hinders the determination of the velocity of the dense shell. We attribute the disappearance of these lines to an increase in the coefficient of opacity in the heated circumstellar matter.

At the same time, several observations made from September 24 to September 28 [30, 31] allow the application of our method.

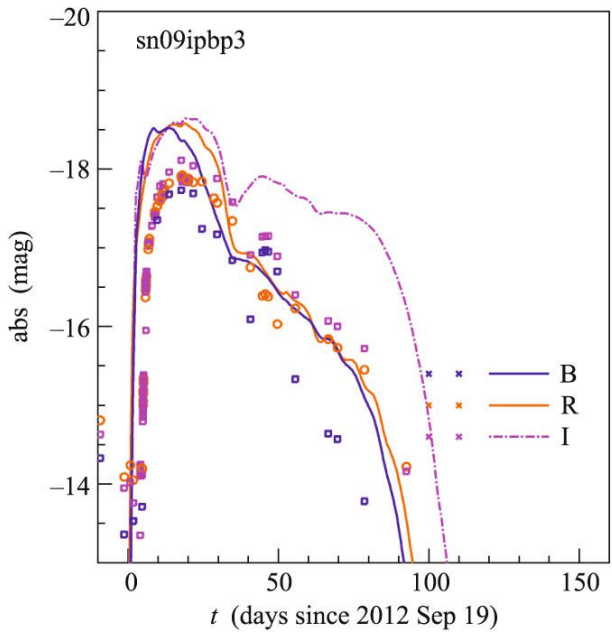

**Fig. 1.** Light curves of the SN 2009ip supernova for the sn09ipbp3 model. The absolute luminosity in the filters marked as B, R, and I versus the observation time in days after the explosion time.

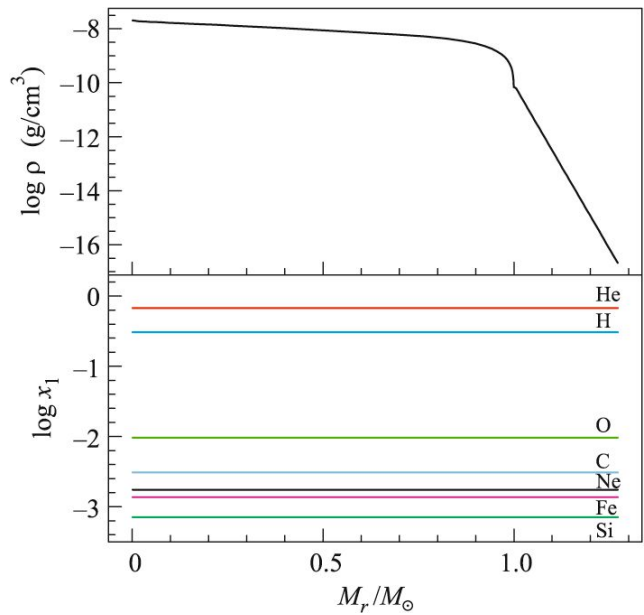

**Fig. 2.** (Upper panel) Initial density distribution and (lower panel) chemical composition of the shell of the supernova for the sn09ipbp3 model.

In [32], we determined the distance to the supernova as $D = 20.1 \pm 0.8$ Mpc in excellent agreement with the known distance modulus to the parent galaxy $\mu = 31.55^m$ [33], which is equivalent to $D = 20.4$ Mpc.

### 2.3. SN 2010jl

Newton and Puckett [34] discovered this supernova on November 3, 2010 (UT) in the UGC5189A galaxy located at a distance of 50 Mpc [35]. The galactic absorption to the supernova is $E_{B-V} = 0.027$, $A_V = 0.083$ [36, 37].

We exemplify how the distance to the supernova can be estimated by the dense shell method from only two observation points without constructing the model of the supernova.

Observations in the stage of an increase of the light curve, when the shock wave moves in the circumstellar matter and forms the dense shell, are necessary. It can be accepted that the temperature $T$ on the photosphere does not change (justification is given below) and the luminosity increases because of an increase in the geometric dimensions of the photosphere moving at the velocity $v_{ph}$.

The broadband photometric data can be obtained from Fig. 1 in [36]. According to these data, in the interval between $t_1 = 10^d$ and $t_2 = 14^d$, the flux in the filter V increased by $0.1^m$ (from $13.7^m$ to $13.6^m$).

According to our estimates, $T = 7000$ K (see Fig. 3 in [36] and Fig. 2 in [35]). Unfortunately, at the neces-

sary times, we have no data on the velocity of the photosphere found from the spectra of broad lines. We used the value $v_{ph} = 5.5 \times 10^8$ cm/s obtained from the approximation of Fig. 12 from [36].

The substitution of the observation data into Eq. (4) gives the distance to the supernova $D = 49$ Mpc in good agreement with the known distance 50 Mpc to the galaxy.

Such an estimate can be obtained by everybody who has observation data. We are going to continue more detailed analysis of the determination of the distance to this supernova. The presented calculation clearly demonstrates that the simple blackbody model can be used to determine the distance if the appropriate time of the evolution of a supernova, namely, the stage of an increase in the luminosity when the shock wave passes in the circumstellar matter cloud surrounding the supernova, is selected.

Below, we consider the applicability of the blackbody model in this case.

## 3. JUSTIFICATION OF THE DENSE SHELL METHOD

We construct the sn09ipbp3 model for the qualitative investigation of the formation of the thin dense shell in the SN 2009ip supernova. The constructed models do not yet reproduce all details of numerous observation data. At the same time, they already qualitatively reproduce the rising part of the light curve of a supernova (Fig. 1), which is most important for the new method.

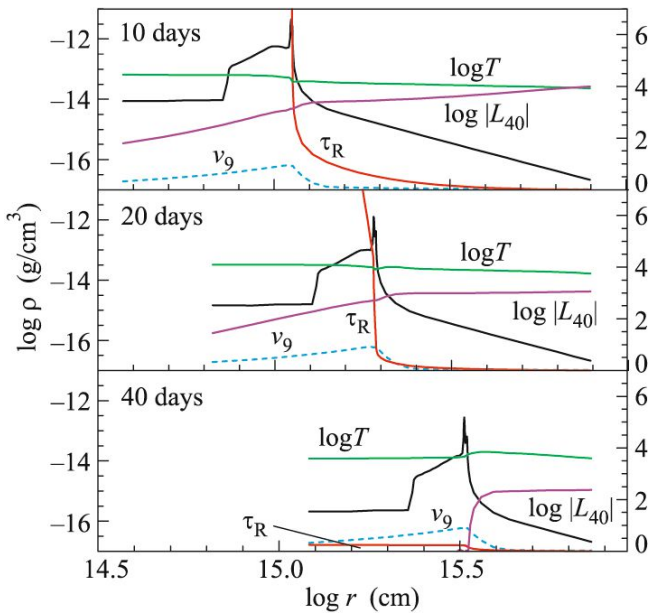

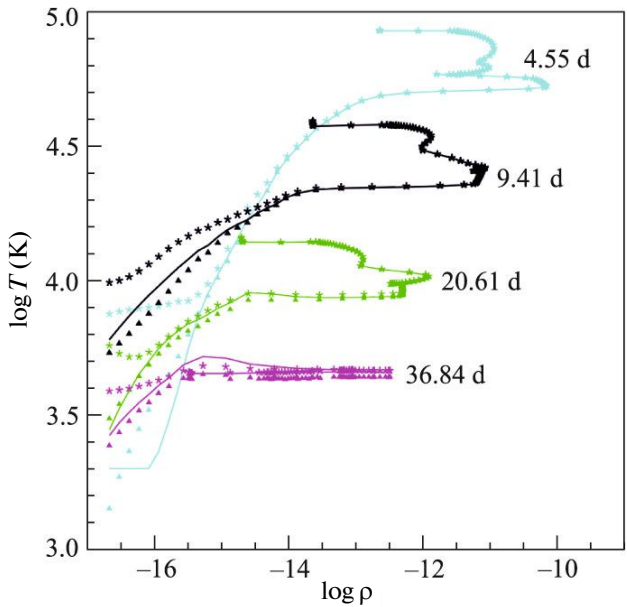

**Fig. 3.** (Color online) (black line) Density $\rho$, (red line) Rosseland optical thickness $\tau$, (green line) temperature $T$, (violet line) luminosity $L$, and (blue line) velocity of the photosphere $v_{ph}$ versus the radius of the supernova for the sn09ipbb3 model at the times $t = 10$, 20, and 40 d. The left ordinate axis shows $\log \rho$ and the right ordinate axis shows $\log \tau$, $\log T$ and normalized $\log L_{40}$, $\log v_9$.

**Fig. 4.** (Solid lines) $\log T$–$\log \rho$ evolution in the local thermodynamic equilibrium approximation. The color ($T_{rad}$) and effective ($T_J$, $J = a T_J^4$) temperatures are marked by asterisks and triangles, respectively. For each given time $t$, each point on the $(T, \rho)$ curve corresponds to a certain Lagrange mass $m$ varying along the curve.

When constructing the model, we solve the transport equations together with the hydrodynamic equations of motion of a gas. We do not impose any constraints on the radiation spectrum $J_\nu$ except for the division of the spectrum into a certain number of groups in frequency in which the radiative parameters (intensity, opaqueness, optical thickness) are independent of the frequency [38, 39]. This makes it possible to realistically and self-consistently describe various complex processes in the shells of supernovae: the propagation of the shock wave and heating of the shell, shock breakout, the passage of the recombination wave through the shell, and luminescence of the supernova in the plateau stage. These calculations were performed in the local thermodynamic equilibrium approximation; i.e., the Saha and Boltzmann formulas are used to calculate the concentrations of ions and population of levels in the equation of state of the plasma in a nonequilibrium radiation field.

An explosion in the sn09ipbb3 is initiated in an extended shell $R = 700 R_\odot$ with the total mass $M = 1.3 M_\odot$ and a uniform distribution of the element (Fig. 2). Such a structure of the shell reproduces the structure of the shell of the presupernova, which discharges external shells shortly before the explosion.

## 3.1. Justification of the Attachment of the Photosphere to the Dense Shell and $T \approx const$

The formation of the thin dense shell can be seen in Fig. 3, where the structure of the shell is shown at the time instants $t = 10$, 20, $40^d$. According to the $\tau(r)$ plot, the main part of the optical thickness is collected just in the thin dense shell. The luminosity $L$ after the passage of the thin dense shell at the time $t = 10^d$ still increases slightly; then, shells in front of the thin dense shell continuously become more transparent and become completely transparent to $t = 20^d$. Thus, the photosphere is completely glued to the thin dense shell. At $t = 40^d$, the entire shell is transparent and light is emitted only from the thin dense shell.

Analysis of Fig. 3 indicates that $T$ is approximately constant and the photosphere moves together with the thin dense shell at the velocity $v_{ph}$ equal to the velocity $u$ of the matter of the thin dense shell. This means that the velocity of the photosphere can be directly obtained by measuring the broadening of spectral lines.

## 3.2. Justification of the Blackbody Model

As was mentioned above, we do not impose constraints on the radiation spectrum $J_\nu$. Therefore, in our models, it can be significantly different from the blackbody spectrum. This can be seen from the curves shown in Fig. 4, which illustrates the time evolution of three characteristic temperatures: (solid line) temperature of matter $T$, (stars) color temperature, and (triangles) effective radiation temperature $T_J$.

It is seen that the blackbody approximation is well applicable inside the shell and all temperatures coincide with each other, whereas radiation near the boundary of the supernova, i.e., at low density, is no longer described by the Planck spectrum and is not in equilibrium with matter. In other words, local thermodynamic equilibrium conditions are broken.

We consider the processes occurring in this case and their effect on the dense shell method and analyze how our equations should be changed in order to take into account the nonequilibrium of radiation in outer layers of the shell of a supernova.

The calculations of the light curves and spectra of supernovae reported in [40–42] involve the solutions of the total system of kinetic equations, which include shock and radiative transitions between atomic levels and continuum for all possible excitation and ionization states of atoms with tens of thousands of levels.

We created the Levels software package for the calculation of the populations of the levels of all atoms and ions in the complete kinetic scheme, where the radiation field is described in the Sobolev approximation. Under the assumption of stationarity, the calculations show that radiative transition rates between levels under our conditions are always noticeably higher than the shock transition rates between the same levels. The latter rates can approach the former ones only for very high lying closely spaced levels, where populations are already insignificant. Furthermore, the photoionization process rate, particularly for hydrogen, beginning with a certain level becomes lower than the shock ionization rate. The local thermodynamic equilibrium approximation for these levels is better applicable because the direct and inverse shock processes are almost completely balanced.

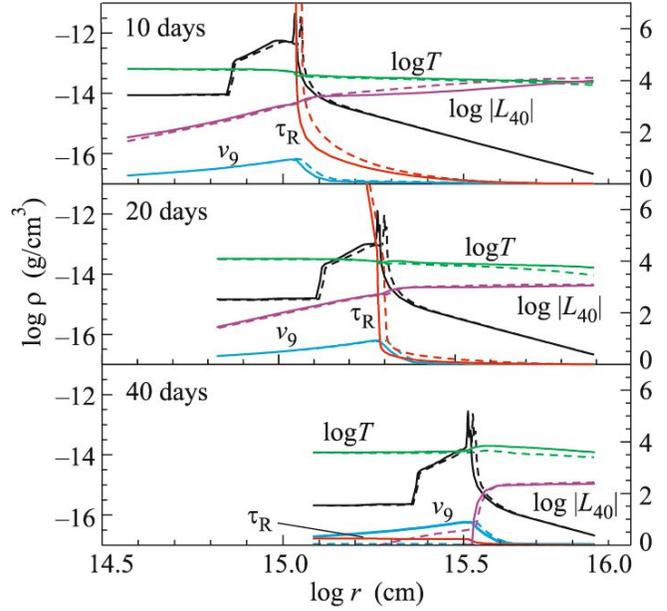

**Fig. 5.** (Color online) (Black line) Density $\rho$, (red line) Rosseland optical thickness $\tau$, (green line) temperature $T$, (violet line) luminosity $L$, and (blue line) velocity of the photosphere according to the calculations in (solid lines) the local thermodynamic equilibrium approximation and (dashed lines) without it for the evolution of the shock wave propagating through the shell of the supernova versus the radius of the supernova. The left ordinate axis shows $\log \rho$ and the right ordinate axis shows $\log \tau$ and $\log T$ normalized to $\log L_{40}$ and $\log v_9$.

The density and temperature in the time interval of interest decrease rapidly to $\rho \leq 10^{-12}$ g/cm$^3$ and $T \leq 1.2 \times 10^4$ K (Fig. 3).

In order to estimate the role of shock processes in our model, we performed the calculation for two cases, with and without the inclusion of shock processes. The results of these calculations are compared in the table.

Concentrations of ions $N_{ijk}$ calculated with and without the inclusion of shock processes for the mixture coinciding in chemical composition with the model case (Fig. 2) at $T = 1.2 \times 10^4$ K and $\rho = 10^{-12}$ g/cm$^3$. The integer part of a number in the leftmost column means the degree of ionization (1 corresponds to neutral) and the fractional part indicates the excitation level (1 corresponds to the ground level)

| H | | C | |
|---|---|---|---|
| 1.01 | 1.7161e+08/1.7735e+09 | 2.01 | 1.5254e+08/1.5254e+08 |
| 1.02 | 5.6416e+03/5.6815e+03 | 2.02 | 2.4570e+05/2.4570e+05 |
| 2.01 | 1.8353e+11/1.8193e+11 | 3.01 | 4.2105e+03/4.2474e+03 |
| He | | Fe | |
| 1.01 | 1.0393e+11/1.0392e+11 | 1.01 | 1.8969e+00/1.8682e+00 |
| 2.01 | 2.5296e+07/2.5517e+07 | 2.01 | 1.7754e+06/1.7636e+06 |
| C | | 2.02 | 1.1765e+06/1.1687e+06 |
| 1.01 | 3.4388e+04/3.4085e+04 | 3.01 | 1.1634e+07/1.1658e+07 |
| 1.02 | 3.4781e+03/3.4475e+03 | | |

It can be seen that shock processes can be neglected under these conditions.

The above properties make it possible to solve the equation of state with the modified nebular approximation [43] for the calculation of the populations of levels and concentrations of ions. The results of such calculations in the local thermodynamic equilibrium approximation and without it are shown in Fig. 5 by the solid and dashed lines, respectively. The calculation without the local thermodynamic equilibrium approximation provides the following conclusions.

—Nonequilibrium radiation in the equation of state enhances ionization in the plasma, making it less transparent. It can be seen that $\tau$ in such a plasma increases toward the center more rapidly.

—An increase in the coefficient of opacity makes a contribution to the radiative term in the equilibrium for $v$. As can be seen in Fig. 5, the shell expansion velocity increases.

—A faster expansion shifts the thin dense shell in radius with respect to the local thermodynamic equilibrium approximation.

The shift of the thin dense shell is small and does not change in time. Hence, it does not affect the results obtained in the dense shell method for which change in $dr$ is important. It can also be seen that the inclusion of processes without local thermodynamic equilibrium hardly changes the luminosity $L$ and temperature $T$. For this reason, we assume that the dense shell method also provides good results for the simple blackbody model, which is confirmed by the above examples.

## 4. CONCLUSIONS

The most important advantage of the dense shell method for the determination of distances to type-IIn supernovae is that it is a direct method that does not require the distance ladder or any preliminary statistical estimates. If the observational data are available, the dense shell method makes it possible to immediately obtain the distance to a supernova, as was shown for the SN 2010jl supernova.

Using the dense shell method, we have obtained the distances to three supernovae: $D \approx 68^{+19}_{-15}$ Mpc to the SN 2006gy supernova, $D \approx 20.1 \pm 0.8$ Mpc to the SN 2009ip supernova, and $D \approx 49$ Mpc to the SN 2010jl supernova.

The formation of the true photosphere in the thin dense shell and the motion of the photosphere together with the thin dense shell allow the use of the blackbody model in the calculations with a tolerable accuracy. Processes without local thermodynamic equilibrium weakly affect the structure of the thin dense shell and, thereby, the estimated distance.

The works reviewed in this paper were supported by the Russian Foundation for Basic Research (project no. 10-02-00249) and in part by the Government of the Russian Federation (contract no. 11.G34.31.0047).

*Translated by R. Tyapaev*